\documentclass{article}
\usepackage{authblk}
\usepackage{url}
\usepackage{braket}
\usepackage{amsmath}
\usepackage{amsthm}
\usepackage{hyperref}
\usepackage{amssymb}
\usepackage{amsfonts}
\usepackage{color}
\usepackage{float}
\usepackage{graphicx}

\theoremstyle{definition}
\newtheorem{thm}{Theorem}[section]
\newtheorem*{thm*}{Theorem}

\newtheorem*{defn*}{Definition}

\newtheorem*{lem*}{Lemma}

\newtheorem*{rem*}{Remark}

\newtheorem*{con*}{Conjecture}

\newtheorem*{cor*}{Corollary}
\newtheorem{prop}[thm]{Proposition}
\newtheorem*{prop*}{Proposition}

\newtheorem*{hypoth*}{Hypothesis}

\newtheorem*{claim*}{Claim}

\title{Foundation of Quantum Optimal Transport and Applications}
\author{Kazuki Ikeda\thanks{kazuki7131@gmail.com}}
\affil{Department of Physics, Osaka University, Toyonaka, Osaka 5600043, Japan}
\date{}

\begin{document}

\maketitle
\begin{abstract}
   Quantum optimal transportation seeks an operator which minimizes the total cost of transporting a quantum state to another state, under some constraints that should be satisfied during transportation. We formulate this issue by extending the Monge-Kantorovich problem, which is a classical optimal transportation theory, and present some applications. As examples, we address quantum walk, quantum automata and quantum games from a viewpoint of optimal transportation. Moreover we explicitly show the folk theorem of the prisoners' dilemma, which claims mutual cooperation can be an equilibrium of the repeated game. A series of examples would show generic and practical advantages of the abstract quantum optimal transportation theory. 
\end{abstract}
\setcounter{tocdepth}{2}
\newpage
\section{Introduction}
Optimization is ubiquitous in various studies. In the modern physics literature, a preferred physical quantum quantity can be obtained by optimizing (minimizing or maximizing) a certain functional. There are various formulations and definitions of optimization problems. In this article we address optimal transportation, which is a problem to find a optimized way of transporting objects. A rigorous mathematical definition of optimal transportation was given by Gaspard Monge in 1781 \cite{monge1781}, and since then a number of authors worked on the problem. From a viewpoint of physics, the conventional optimal transportation theory basically is established in a classical manner. Therefore recent advances in quantum technologies endow us with motivation to address a quantum version of Monge's problem, that is "What is an operator which minimizes the total cost of transporting a quantum state to another state?". As shown later, this problem can be formulated in various ways, depending on constraints imposed on a transportation process. Optimal transportation seeks an optimized map or operator of transportation by a non-perturbative way. This is in contrast to modern physical method of deriving a classical saddle. However this conventional method, in general, does not always become the best way to investigate non-local physics. Hence work presented here has a potential advantage when one considers an approach to global quantum physics. In this note, we present some formulations of quantum optimal transportation and apply them to some practical examples.  

Attempt of solving not necessarily physical problems by a physical method is increasingly gaining a lot of interest, due to recent progress in quantum computers and quantum information technology. In fact, solving combinatorial optimization problem by a quantum physics way is known as quantum annealing \cite{PhysRevE.58.5355}, which is partly implemented with superconducting qubits \cite{Johnson2011} and applied to some NP-hard problems \cite{ikeda2019application}. Indeed, we later show that some optimal transportation problems can be implemented by a formalism of quantum annealing. While a quantum annealing can solve only discrete problems, in this article, we address problems with uncountable number of degrees of freedom. Though such a computer that solves non-discrete problems have not existed so far, our formulation of continuous quantum Monge problems can be useful when a machine with an ultimate computational capability is realized. Indeed quantum field theory is quantum mechanics with uncountable number of degrees of freedom and nature implements quantum field theoretical algorithm by a yet-unknown way.

This piece is orchestrated as follows. In section 2, we first give a brief review on the Monge-Kantorovich optimal transportation theory and give various quantum extensions. Then we address some examples based on our formalism. Especially we describe a generic relation among quantum walk, quantum automata and quantum games in a context of optimal transportation. Game theory plays a fundamental role in a fairly large part of the modern economics. Optimal transportation seeks an economically best way of transportation, hence it will be natural to ask how it can be useful to game theory. In this work, we investigate a repeated quantum in terms of quantum optimal transportation and show the folk theorem of the repeated quantum game, that is there exists an equilibrium strategy of the repeated game. Various formulations of quantum games are proposed by many authors, whereas less is known for repeated games. Especially, this is the first work that investigate an infinitely repeated quantum game and the folk theorem. 

\section{Quantum Optimal Transport}\label{sec:opt}
\subsection{The Monge-Kantorovich Problem}
 \begin{figure}[H]
\centering
\includegraphics[width=5cm]{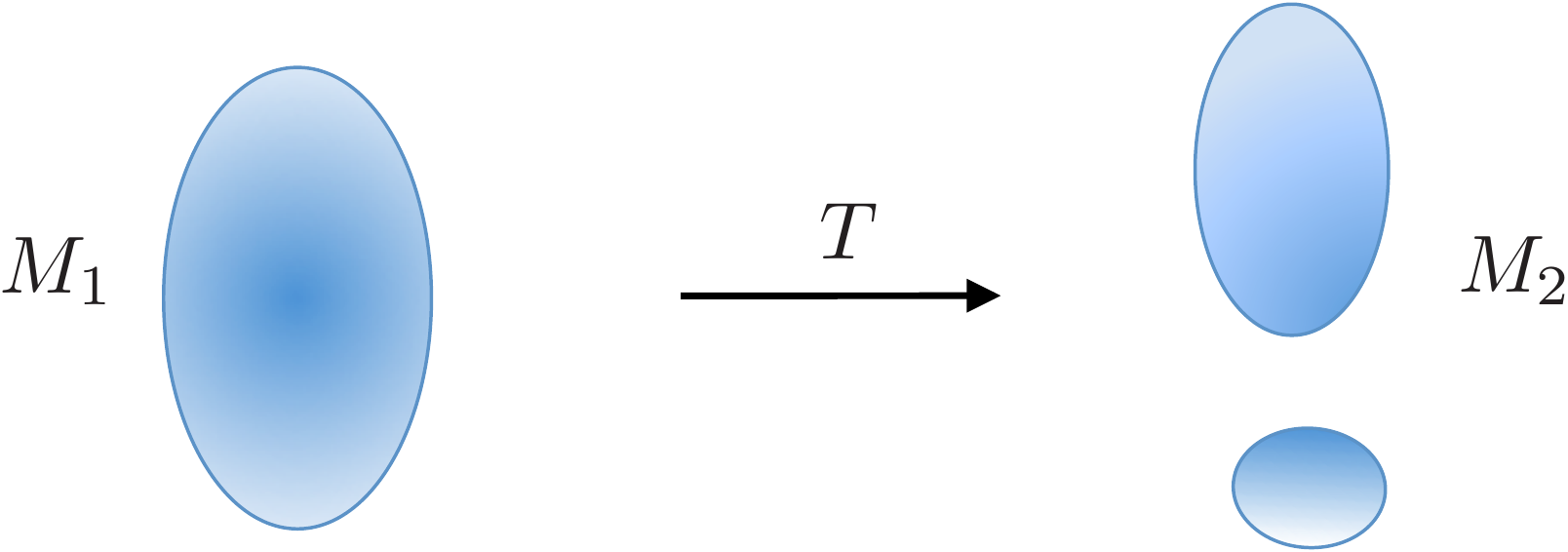}
\end{figure}
 We consider the Monge-Kantorovich problem \cite{monge1781,k1942}. Let $(X,\mu), (Y,\nu)$ be two probability spaces. The original Monge problem is to find a bijective map $T:X\to Y$ which minimizes the total cost
 \begin{equation}\label{op:Monge}
     C(T)=\int_{X}c(x,T(x))\mu(dx),   
 \end{equation}  
 where $c(x,T(x))$ is some function on $X$. We write the transported distribution as $\nu=T_\#\mu$. Monge's problem is reformulated by Kantorovich in such a way that one finds an optimal plan $\pi=(1\times T)_\#\mu$, which satisfies  $\pi(A,Y)=\mu(A), \pi(X,B)=\nu(B)$ for all measurable sets $A\subset X,B\subset Y$ and 
 \begin{equation}
     \int_{X\times Y}c(x,y)\pi(dxdy)=\int_{X}c(x,T(x))\mu(dx). 
 \end{equation}
Hence the optimized transportation plan is $\pi(dxdy)=\mu(dx)\delta_{T(x)}(dy)$. Let $(T_t(x))_{t\in[0,1]}$ be associated optimal flow and $(\mu_t)_{t\in [0,1}$ be a family of curves $\mu_t=T_t{}_\#\mu$. Then they naturally obey the equation of continuity
\begin{equation}\label{eq:eoc}
    \partial_t\mu+\nabla(v\mu)=0, 
\end{equation}
where $v(t,T_t(x))=\frac{d}{dt}T_t(x)$. 

Now let us formulate the Monge-Kantorovich problem with a Hamiltonian formalism. Before moving to quantum cases, we first consider the  classical problem. Let $q: X\times Y\to [0,1]$ be a function by which $\mu_0(x)q(x,y)$ indicates the amount transported to $y$ from $x$. Then a solution of classical Monge-Kantrovich problem is a ground state of the following Hamiltonian:
\begin{align}\label{eq:otp}
\begin{aligned}
    H=&\int dxdyc(x,y)\mu(x)q(x,y)+\int dx\left(\int dy\mu(x)q(x,y)-\mu(x)\right)^2\\
    &+\int dy\left(\int dx\mu(x)q(x,y)-\nu(y)\right)^2. 
\end{aligned}
\end{align}
The first term is the cost of transportation, the second term is the penalty term which requires for any $x$ the sum of the transported amount from $x$ to $y$ is equal to the sum of the amount $\mu(x)$ at $x$, and the third term implies that the required amount $\nu(y)$ should be delivered to all $y$ without loss. The problem is solved by finding $\{q(x,y)\}_{(x,y)\in X\times Y}$ that minimizes the Hamiltonian \eqref{eq:otp}. 

In practice, we can numerically simulate the problem in a discrete situation, by finding the ground state of the classical Hamiltonian 
\begin{align}\label{eq:otp2}
\begin{aligned}
    H=&\sum_{x,y}c(x,y)\mu(x)q(x,y)+\sum_{x}\left(\sum_{y}\mu(x)q(x,y)-\mu(x)\right)^2\\
    &+\sum_{y}\left(\sum_{x}\mu(x)q(x,y)-\nu(y)\right)^2.
\end{aligned}
\end{align}
If $q$ takes value in $\{0,1\}$, this Hamiltonian \eqref{eq:otp2} works for quantum annealing. This corresponds to the Hamiltonian of the Hitchcock transportation problem. 

\subsection{Quantum Optimal Transport}
In what follows we work on a Euclidean space $X=Y=\mathbb{R}^d$. Occasionally we use $X$ or $Y$ to emphasize a reference space and a target space. We may define the quantum optimal transport by saying that find an operation $T:\mathcal{H}(X)\to\mathcal{H}(Y)$ which minimizes the total cost (or maximizes the total reward) when a given wave function $\psi_0(x)=\langle{x}|\psi_0\rangle$ on $M$ is transported to another $\psi_1(y)=\langle{y}|\psi_1\rangle$ on $N$. So we formulate the problem by the following functional $I[T]$ of $T$ defined by 
\begin{align}\label{eq:v0}
\begin{aligned}
    I[T]=&\int_Y dy\bigg|\int_X dx\sqrt{c(x,y)}\bra{y}T\ket{x}\langle{x}|\psi_0\rangle\bigg|^2\\
    &+\int_Y dy\lambda(y)\left( \bra{y}T|\psi_0\rangle-\langle y|\psi_1\rangle\right),
\end{aligned}
\end{align}
where $\lambda(y)$ is a Lagrange multiplier. We expand 
\begin{equation}
    T\ket{x}=\int_YdyT(x,y)\ket{y},
\end{equation}
where $T(x,y)\in\mathbb{C}$ should satisfy the unitarity condition
\begin{equation}
    \int_Ydy|T(x,y)|^2=1,~~\forall x\in X. 
\end{equation}
Alternatively we can do the same business by introducing an operator $\widehat{CT}$ which acts on $\ket{x}$ as 
\begin{equation}
    \widehat{CT}\ket{x}=\int_Ydy'\sqrt{c(x,y')}T(x,y')\ket{y'} 
\end{equation}
and redefine the cost term with $\bra{y}\widehat{CT}\ket{x}$, which is equivalent to $\sqrt{c(x,y)}\bra{y}T\ket{x}$. In fact the following formula holds.  
\begin{align}
    \bigg\|\int_Xdx\widehat{CT}\ket{x}\bra{x}\psi_0\rangle\bigg\|^2=\int_Ydy\bigg|\int_Xdx\langle{y}|\widehat{CT}\ket{x}\bra{x}{\psi_0}\rangle\bigg|^2. 
\end{align}
Proof is simple. The L.H.S. is 
\begin{align}
    &\int_X{dxdx'}\left(\bra{\psi_0}x'\rangle\bra{x'}\widehat{CT}^\dagger\right)\left(\widehat{CT}\ket{x}\bra{x}\psi_0\rangle\right)\\\label{eq:quantum}
    =&\int_X{dxdx'}\int_Y{dydy'}\bra{\psi_0}x'\rangle\bra{x'}\widehat{CT}^\dagger\ket{y'}\bra{y'}y\rangle\bra{y}\widehat{CT}\ket{x}\bra{x}\psi_0\rangle\\
    =&\int_Ydy\bigg|\int_Xdx\langle{y}|\widehat{CT}\ket{x}\bra{x}{\psi_0}\rangle\bigg|^2
\end{align}
 So we can interpret the state $\int_Xdx\widehat{CT}\ket{x}\bra{x}\psi_0\rangle$ as the quantum version of Monge's integral \eqref{op:Monge}. $\big\|\int_Xdx\widehat{CT}\ket{x}\psi_0\rangle\big\|^2$ gives amplitude of states after transported. In general, the cost $\sqrt{c(x,y)}$ would make a transition process non-unitary, as evolution of a particle interacting with a heat bath. We may restrict to the case $0\le|c(x,y)|\le1$ for all $x\in X,y\in Y$ and consider the problem $\sup_{\widehat{T}}I[\widehat{T}]$, instead of $\inf_{\widehat{T}}I[\widehat{T}]$. If $c(x,y)=1$ everywhere, any $\widehat{T}$ which realizes $\ket{\psi_1}=\widehat{T}\ket{\psi_0}$ can be a solution of the problem. 

Our framework can address the classical Monge's problem as well. Using the formula
\begin{align}
\begin{aligned}
    \big\|\widehat{CT}\ket{x}\big\|^2&=\int_Ydydy'\sqrt{c(x,y)c(x,y')^*}T(x,y)T(x,y')^*\bra{y}{y'}\rangle\\
    &=\int_Ydyc(x,y)|T(x,y)|^2,
\end{aligned}
\end{align}
we find that the functional of $\widehat{T}$
\begin{align}
    I_1[\widehat{T}]&=\int_Xdx\big\|\widehat{CT}\ket{x}\bra{x}\psi_0\rangle\big\|^2\\
    &=\int_Xdx\int_Ydyc(x,y)|T(x,y)|^2\mu(x)
\end{align}
describes the classical Monge's optimal transportation. Here $|\bra{y}\widehat{T}\ket{x}|^2=|T(x,y)|^2$ plays the role of $q(x,y)$. Therefore the classical formulation of the problem with a strict constraint on quantum states is
\begin{align}
\label{eq:cH}
    I[\widehat{T}]=&\int_Y dy\int_Xdx\bigg| \sqrt{c(x,y)}\bra{y}\widehat{T}\ket{x}\langle{x}|\psi_0\rangle\bigg|^2\\
    &\label{H:y}+\int_Y dy\lambda_1(y)\left( \bra{y}\widehat{T}|\psi_0\rangle-\langle y|\psi_1\rangle\right),
\end{align}

This functional can be also obtained when quantum interaction between two different positions is lost, namely $\langle x\ket{\psi_0}\bra{\psi_0}x'\rangle=|\bra{x}\psi_0\rangle|^2\delta(x-x')$ holds in the equation \eqref{eq:quantum}. 

\paragraph{Dynamical Approach}
Now let us consider a dynamical approach. Let $(\widehat{T_t})_{t\in[0,1]}$ be a
family of operators $\widehat{T_t}:\mathcal{H}(X)\to \mathcal{H}(X)$ defined by 
\begin{align}
    \widehat{T_t}\ket{x}=\int_X dx'T_t(x,x')\ket{x'},
\end{align}
where $T_t(x,x')\in\mathbb{C}$ satisfies $\int_Xdx'|T_t(x,x')|^2=1$ for any $x$ and $t$. With respect to this $\widehat{T_t}$ we consider a family $(\widehat{CT_t})_{t\in[0,1]}$ of operators with cost 
\begin{equation}
    \widehat{CT_t}\ket{x}=\int_X dx'c(x,x')T_t(x,x')\ket{x'}. 
\end{equation}
Suppose $T_t(x,x')$ and $c(x,x')$ are smooth and finite with respect to any choice of parameters. The problem is to find $\widehat{T}=(\widehat{T_t})_{t\in[0,1]}$ which satisfies  $\ket{\psi_1}=\widehat{T_1}\ket{\psi}$ and minimizes the total cost (or maximizes the total reward) for the quantum case 
\begin{equation}
    \int_0^1 dt \bigg\|\int_Xdx\widehat{CT_t}\ket{x}\bra{x}\psi_0\rangle\bigg\|^2
\end{equation}
and for the classical case 
\begin{equation}
    \int_0^1 dt \int_Xdx\big\|\widehat{CT_t}\ket{x}\bra{x}\psi_0\rangle\big\|^2. 
\end{equation}

The classical formulation is obtained by another way. he information entropy of a quantum system is expressed with the
density operator $\rho$ in such a way that 
\begin{equation}
    S(\rho)=-\text{Tr}(\rho\log\rho). 
\end{equation}
We define $\rho_t(x)=T_t\ket{x}\bra{x}T_t^\dagger$ and the trace operation by 
\begin{equation}
    \text{Tr}\rho^T_t(x)=\int_Y dy\bra{y}\rho_t(x)\ket{y}. 
\end{equation}
By definition, $\text{Tr}\rho^T_t(x)=\int_Xdy|T_t(x,y)|^2$, which is equal to $1$ due to the unitarity and have the conservation law  
\begin{equation}
    \frac{d}{dt}\text{Tr}\rho^T_t(x)=0,~~\forall x\in X. 
\end{equation}

Similarly, we define 
\begin{equation}
    \rho^{CT}_t(x)=\widehat{CT_t}\ket{x}\bra{x}\widehat{CT_t}^\dagger, 
\end{equation}
whose trace $\text{Tr}\rho^{CT}_t(x)=\int_Ydy|\sqrt{c(x,y)}T_t(x,y)|^2$. Using this, we can write the functional $I_t[T]$ in a simple form
\begin{equation}
    I_t[T]=\int_X\rho^{CT}_t(x)\mu(x)dx. 
\end{equation}
Moreover 
\begin{equation}
    \rho^{\psi_0}_t(x)=\widehat{T_t}\ket{x}\bra{x}\psi_0\rangle\langle{\psi_0}\ket{x}\bra{x}\widehat{T_t}^\dagger
\end{equation}
gives the total amount 
\begin{equation}
    \text{Tr}\rho_t^{\psi_0}(x)=\int_Ydy|T_t(x,y)|^2\mu(x)
\end{equation}
transported to $y$ from $x$ at $t$. The unitarity requires 
\begin{equation}
    \text{Tr}\rho^\psi_t(x)=\mu(x). 
\end{equation}

We may write 
\begin{equation}
    \mu_t(y)=\int_Xdx|T_t(x,y)|^2\mu(x). 
\end{equation}
Then at the end of transportation $t=1$, the density precisely obeys  $\mu_1(y)=|\langle y|\psi_1\rangle|^2=\nu(y)$, which agrees with the constraint on $\langle y|T_1|\psi_0\rangle=\bra{y}\psi\rangle$. In this way we can recover the classical picture of optimal transportation. 

\if{
\paragraph{Entanglement}
Let $\ket{\Psi}$ be the following entangled state consisted of initial state and the transported state
\begin{align}
\begin{aligned}
    \ket{\Psi}&=\int_ Mdx\ket{x}\otimes T\ket{x}\\
    &=\int_Mdx\int_NdyT(x,y)\ket{x}\otimes\ket{y}. 
\end{aligned}
\end{align}
With respect to this $\ket{\Psi}$, the density operator on $N$ is obtained by integrating out $M$'s degree of freedom
\begin{align}
\rho_N&=\int_Mdx\bra{x}\Psi\rangle\langle\Psi\ket{x}\\
&=\int_Ndydy'T(x,y)T(x,y')^*\ket{y'}\bra{y}. 
\end{align}

We consider $N=M$ and suppose a family $(T_t)_{t\in[0,1]}$ of unitary operators. Spectrum theorem... 
}\fi
 

\subsubsection{Variant 1}
So far we have discussed the case where initial state is transformed into a promised state. In practise, a quantum state is not a physical observable and this constraint is too hard, thereby it would be better to work with relaxed constraints. One of the most practical requirements is that quantum states are efficiently transported so that the total cost is as small as possible and the transported quantum state forms an expected probability distribution.
The corresponding functional is defined with a Lagrange multiplier $\lambda(y)$ in such a way that
\begin{align}\label{eq:v1}
\begin{aligned}
    I[\widehat{T}]&=\int dy\bigg| \int dx\sqrt{c(x,y)}\langle y|\widehat{T}|x\rangle\langle x|\psi_0\rangle\bigg|^2\\
    &+\int dy\lambda(y)(\mu(y)-|\langle y|\widehat{T}|\psi_0\rangle|^2). 
\end{aligned}
\end{align}
When two the observable $\mu(x)=|\psi_0(x)|^2$ and $\nu(y)=|\psi_1(y)|^2$ are given, find a unitary operation which minimizes the cost of transporting wave function. With respect to a wave function $\psi:\mathbb{R}\to\mathbb{C}$, we define its support $\text{supp}(\psi)$ by 
\begin{equation}
    \text{supp}(\psi)=\{x\in \mathbb{R}:\psi(x)\neq0\}. 
\end{equation}
Let $\psi_a(x),\psi_b(x)$ be normalized wave functions on $X$. We write $\psi_{a\pm b}(x)=\frac{1}{\sqrt{2}}(\psi_a(x)\pm\psi_b(x))$. Density distributions $|\psi_{a+b}(x)|^2$ and $|\psi_{a-b}(x)|^2$ become equal to each other if they are not correlated $\text{supp}(\psi_{a+b})\cap\text{supp}(\psi_{a-b})=\emptyset$. While the original functional \eqref{eq:v0} requires a coincidence between a mapped state and a target state, the functional \eqref{eq:v1} only demands a coincidence between a mapped distribution and a target distribution. In this sense, the functional \eqref{eq:v1} looks practical. 

Moreover it is also possible to work with the classical formulation of the problem, by optimizing the functional 
\begin{align}
\begin{aligned}
I[\widehat{T}]&=\int dy \int dx\bigg|\sqrt{c(x,y)}\langle y|\widehat{T}|x\rangle\langle x|\psi_0\rangle\bigg|^2\\
    &+\int dy\lambda(y)(\nu(y)-|\langle y|\widehat{T}|\psi_0\rangle|^2)
\end{aligned}
\end{align}
This functional corresponds to solving the classical problem by a quantum method. Generally it would be hard to find the optimized $\widehat{T}$. 


\subsubsection{Variant 2}
We consider quantum optimal transport $\widehat{T}$ so that the transported states becomes as close as possible to the desired state in along with minimizing the total transportation cost. Instead of using a Lagrange multiplier, we consider fidelity $F(\psi_1,\widehat{T}\psi_0)=\frac{|\langle\psi_1|\widehat{T}|\psi_0\rangle|^2}{\|\widehat{T}\ket{\psi_0}\|^2}$ to measure the quantum distance $D(\psi_1,\widehat{T}\psi_0)=1-F(\psi_1,\widehat{T}\psi_0)$ between two states.    
\begin{equation}
    I[\widehat{T}]=\int_Ydy\bigg|\int_X dxc(x,y)\bra{y}\widehat{T}\ket{x}\langle{x}|\psi_0\rangle \bigg|^2 +D(\psi_1,\widehat{T}\psi_0). 
\end{equation}

\subsubsection{Variant 3}
Let $\ket{\psi_0}\in\mathcal{H}(X)$ and $\ket{\psi_1}\in\mathcal{H}(Y)$ be given states. Let $\widehat{C}:\mathcal{H}(X)\to\mathcal{H}(Y)$ be a given cost operator, and $\widehat{T}:\mathcal {H}(X)\to\mathcal{H}(X)$ be a unitary operator. We define the problem to find an optimal $\widehat{T}$ which maximize the following functional 
\begin{equation}
    I[\widehat{T}]=|\bra{\psi_1}\widehat{C}\widehat{T}\ket{\psi_0}|^2. 
\end{equation}
This would be understood as the problem to find a operator which maximize the probability amplitude of sending an initial state $\ket{\psi_0}$ to a finial state $\ket{\psi_1}$ with the cost $\widehat{C}$. By inserting $\int_X\ket{x}\bra{x}=1$ and $\int_Y\ket{y}\bra{y}=1$, the problem is equivalent to   
\begin{align}\label{eq:V6}
    I[\widehat{T}]=\bigg|\int_Ydy\int_Xdx\bra{\psi_1}y\rangle\bra{y}\widehat{C}\ket{x}\bra{x}\widehat{T}\ket{\psi_0}\bigg|^2. 
\end{align}
The cost $\bra{y}\widehat{C}\ket{x}$ would correspond to $\sqrt{c(x,y)}$ in the previous cases.

It is also possible to consider the cost to obtain a given final state $\ket{\psi_1}\in\mathcal{H}(Y)$ by integrating all possible initial states $\ket{\psi_\lambda}\in\mathcal{H}(X)~(\lambda\in[0,1])$, and satisfies 
\begin{equation}
    \int_0^1 d\lambda\ket{\psi_\lambda}\bra{\psi_\lambda}=1. 
\end{equation}
Here we assume that for any $\ket{\psi}\in\mathcal{H}(X)$, there is some $\lambda$ and $\ket{\psi}=\ket{\psi_\lambda}$. The cost to obtain $\ket{\psi_1}\in\mathcal{H}(Y)$ is 
\begin{equation}
    I[\widehat{T}]=\int_0^1 d\lambda|\bra{\psi_1}\widehat{C}\widehat{T}\ket{\psi_\lambda}|^2
\end{equation}
and the problem is to find $T$ which minimizes it. Similarly we can evaluate the cost to start with a given initial state $\ket{\psi_0}\in\mathcal{H}(X)$, by integrating out all final states 
\begin{equation}\label{eq:v4}
    I[\widehat{T}]=\int_0^1 d\lambda|\bra{\psi_{\lambda}}\widehat{C}\widehat{T}\ket{\psi_0}|^2.
\end{equation}

\subsubsection{Variant 4}
Initiated by the functional \eqref{eq:V6}, we formulate a functional by
\begin{equation}
    I[\widehat{T}]=\int_ Ydy\bigg| \int_Xdx\bra{\psi_1}y\rangle\bra{y}\widehat{C}\ket{x}\bra{x}\widehat{T}\ket{\psi_0}\bigg|^2 
\end{equation}
and the classical case by
\begin{equation}
    I[\widehat{T}]=\int_Ydy\int_Xdx\bigg| \bra{\psi_1}y\rangle\bra{y}\widehat{C}\ket{x}\bra{x}\widehat{U}\ket{\psi_0}\bigg|^2. 
\end{equation}
It requires a further investigation to unveil how they could be useful to optimal transportation or to any physical system.

\subsubsection{Variant 5}
Another variation of this game is given by operator $\widehat{CT}$ and maximizing the reward
\begin{equation}\label{eq:amp}
    I[\widehat{T}]=\big|\bra{\psi_1}\widehat{CT}\ket{\psi_0}\big|^2. 
\end{equation}
 By inserting $\int_Ydy\ket{y}\bra{y}=1$ and $\int_Xdx\ket{x}\bra{x}=1$, the functional is 
\begin{align}
\begin{aligned}
I[\widehat{T}]&=\bigg|\int_Ydy\int_Xdx \bra{\psi_1}y\rangle \bra{y}\widehat{CT}\ket{x}\langle x\ket{\psi_0} \bigg|^2      \\
&=\bigg|\int_Ydy\int_Xdx \sqrt{c(x,y)} \bra{\psi_1}y\rangle \bra{y}\widehat{T}\ket{x}\langle x\ket{\psi_0} \bigg|^2
\end{aligned}
\end{align}
The functional \eqref{eq:amp} is the transition amplitude of non-unitary scattering process. By integrating out the degree of freedom about initial or final states, we obtain the cost of ending with $\ket{\psi_1}$ or starting with $\ket{\psi_0}$, respectively.   

A way to introduce dynamics into our model is to consider the following functional of a family $\widehat{T}=\{\widehat{T_t}\}_{t\in[0,1]}$ of transportation operators: 
\begin{equation}\label{eq:aaaa}
    I[\widehat{T}]=\int_0^1 dt \big|\bra{\psi_1}\widehat{CT_t}\ket{\psi_0}\big|^2. 
\end{equation}
This describes the total cost of transporting $\ket{\psi_0}$ to $\ket{\psi_1}$.

\section{Applications of Quantum Optimal Transportation}
\subsection{Costly Quantum Walk}
We consider discrete, one-dimensional and two-state quantum walk  
\begin{equation}
    \ket{\psi_t(x)}=\psi^L_t(x)\ket{L}+\psi^R_t(x)\ket{R}\in\mathbb{C}^2,
\end{equation}
where $\ket{L}=\binom{1}{0},\ket{R}=\binom{0}{1}$ and $\psi_t(x)$ satisfy $\sum_{x=-t}^t\|\psi_t(x)\|^2=1$ for all $t$. Time-evolution $\psi_{t+1}(x)=U_t\psi_t(x)$ of a quantum walker is defined by a unitary matrix $U_t$ in such a way that
\begin{equation}
    \ket{\psi_{t+1}(x)}=U_{L,t}\psi_t(x+1)+U_{R,t}\psi_t(x-1), 
\end{equation}
where $U_{R,t}+U_{L,t}=U_t$ is a two-by-two unitary matrix and $\psi_t(x)$ is a state on $x$ at $t$. More explicitly, a state can be written as 
\begin{align}
    \psi^R_{t+1}(x)=a_t\psi^L_{t}(x+1)+b_t\psi^R_{t}(x+1)\\
    \psi^L_{t+1}(x)=c_t\psi^L_{t}(x-1)+d_t\psi^R_{t}(x-1), 
\end{align}
where $a_t,b_t,c_t,d_t$ are components of a unitary matrix 
\begin{align}
\begin{aligned}
    U_t=
    \begin{pmatrix}
    a_t&b_t\\
    c_t&d_t
    \end{pmatrix},
    ~~U_tU^\dagger_t=1\\
    U_{L,t}=
    \begin{pmatrix}
    a_t&0\\
    c_t&0
    \end{pmatrix},
    U_{R,t}=
    \begin{pmatrix}
    0&b_t\\
    0&d_t
    \end{pmatrix}
\end{aligned}
\end{align}

Suppose the cost of transporting from $x$ to $y$ is given by 
\begin{equation}
    c(x,y)=y^2-x^2. 
\end{equation}
Then the cost operator $\widehat{CU_t}$ acts on $\psi_{t}(x)$ as 
\begin{align}
    \widehat{CU_t}\psi_{t}(x)=
    \begin{pmatrix}
    (2x+1)\left(a_t\psi^L_{t}(x+1)+b_t\psi^R_{t}(x+1)\right)\\
   (-2x+1)\left(c_t\psi^L_{t}(x-1)+d_t\psi^R_{t}(x-1)\right)
    \end{pmatrix}
\end{align}
So the problem is to find best choice of a family $\{U_t\}$ that minimizes the total cost $\sum_{t}\|\widehat{CU_t}\psi_t\|^2$ of transporting an initial state, say, $\psi_0(x)=\delta(x)\ket{R}$ to a given target state or a target distribution.

\subsection{Quantum Cellular Automata}
A two-way quantum finite automaton (2QFA) \cite{Meyer1996} is defined by a 6-tuple 
\begin{equation}\label{automata}
    M=(Q,\Sigma,\delta,q_0,Q_\text{acc},Q_\text{rej}), 
\end{equation}
where $Q,q_0,Q_\text{acc},Q_\text{rej}$ are a set of states, an initial state, a set of accepted states, and a set of rejected states. $\delta:Q\times \Sigma\times Q\times\mathbb{Z}\to\mathbb{C}$ gives a transition amplitude, namely $\delta(q,a,q',D)=\alpha$ means that, when the machine in a state $q$ reads an input letter $a$, the transition amplitude of the state into another $q'$ with a head moving to $D\in\mathbb{Z}$ is $\alpha$. We denote by $\ket{q,x}$ a sate of the machine with the head at $x\in\mathbb{Z}$. So for a given input $a$, a unitary time-evolution $U^a$ of a given state $\ket{q,x}$ is expressed as 
\begin{equation}\label{eq:automata}
    U^a\ket{q,x}=\sum_{q'\in{Q},D\in E(x)}\delta(q,a(x),q',D)\ket{q',x+D},
\end{equation}
where $E(x)\subset\mathbb{Z}$ is a set of lattice vectors defining local neighborhoods for the automaton and $a(x)$ is $a$'s $x$th input letter. Let we define some sets by $\mathcal{H}_\text{acc}=\text{span}\{\ket{q,x}:\ket{q}\in Q_\text{acc}\}$, $\mathcal{H}_\text{rej}=\text{span}\{\ket{q,x}:\ket{q}\in Q_\text{rej}\}$ and $\mathcal{H}_\text{non}=\text{span}\{\ket{q,x}:\ket{q}\in Q\setminus (Q_\text{acc}\cup Q_{rej}\}$. Then the whole Hilbert space is 
\begin{equation}
    \mathcal{H}=\mathcal{H}_\text{acc}\oplus\mathcal{H}_\text{rej}\oplus\mathcal{H}_\text{non}. 
\end{equation}
Let $\pi_\star:\mathcal{H}\to \mathcal{H}_\star,~(\star=\text{acc,rej,non})$ be the natural projections. We denote by $\ket{\Psi_0}=\ket{q_0,0}$ be an initial state. 2QFA works as follows. (1) Pick up $U_t=U^x$ and operate it to
$\ket{\Psi_t}$. We write $\ket{\Psi_{t+1}}=U^x\ket{\Psi_t}$. (2) Measure $\ket{\Psi_{t+1}}$ using projection operators, which makes the state shrink to $\frac{1}{\|\pi_\star\ket{\Psi_{t+1}}\|}\pi_\star\ket{\Psi_{t+1}}$. Further processing halts when either accept or reject is output. The dynamics of a quantum cellular automaton can be described by quantum walk. A way of introducing the cost into this study is to define it as the total steps needed for a state to be accepted or rejected. Suppose dimension of $\mathcal{H}_\text{acc}\oplus\mathcal{H}_\text{rej}$ is finite $n$ and let $\{\psi_i\}_{i=1,\cdots,n}$ be an orthonormal basis of $\mathcal{H}_\text{acc}\oplus\mathcal{H}_\text{rej}$.  Cost arises while states in $\mathcal{H}_\text{non}$ are observed. Therefore a way of defining the total cost that arises until processing halts at certain $t=\tau\in\mathbb{Z}$ is 
\begin{equation}\label{eq:tau}
    \tau[U]=\sum_{t=0}^\tau\sum_{i=1}^n  \Delta\left({\bra{\psi_i}M|\Psi_t\rangle}\right),
\end{equation}
 where $\Delta:\mathbb{C}\ni z\mapsto \Delta(z)\in\{0,1\}$ is non-zero only at $z=0$ and $M$ is the measurement operator. The cost \eqref{eq:tau} is actually a functional $\tau[U]$ of $U=\{U_t\}_{t= 0,\cdots,\tau}$ therefore the problem is to find $U$ which minimizes $\tau[U]$. The complexity of this decision problem should be defined by $\min\tau[U]$.


\subsection{Automata and Games}
\begin{figure}
    \centering
    \includegraphics[width=7cm]{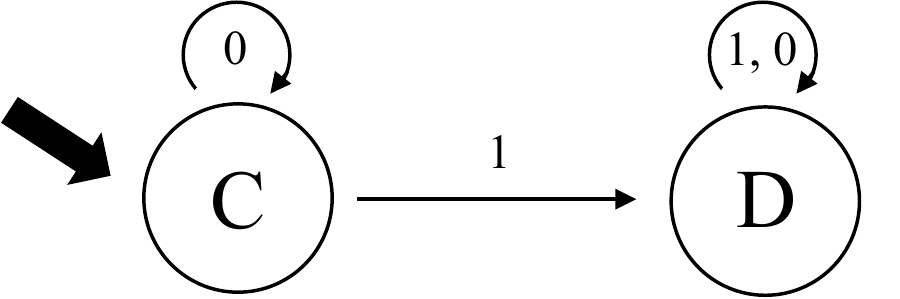}
    \caption{Strategy profile/state transition diagram of the grim trigger strategy. The bold arrow stands for an initial strategy/state.}
    \label{fig:gt}
\end{figure}

Automata and games are widely studied mostly from a viewpoint of computer games. Conway's life game is a well-known example. But here we like to explore a relation with automata and game theory in economics. Especially we are interested in repeated games. Let us first look at classical cases. We prepare a set $\Sigma=\{0,1\}$ of input letters, a set $Q=\{C,D\}$ of states, and a set $Q_\text{acc}\subset\{C,D\}$ of acceptable states. A classical automata consists of those sets, an initial state $q_0$ and a transition function $\delta:Q\times \Sigma\to Q$, which corresponds to a strategy of a game. $\Sigma$ corresponds to a set of signals of opponent's strategy. Signals are updated every round by detecting new ones. As a simple example we consider a two-person prisoners' dilemma (PD) with monitoring where payoff for each agent is given in Table.\ref{tab:prof1}.
\begin{table}[]
    \centering
    \begin{tabular}{c|c|c}
         & C&D \\\hline
        C &(X,X) &(X-Z,X+Y)\\
        D&(X+Y,X-Z)&(X-Z+Y,X-Z+Y)
    \end{tabular}
    \caption{Rewards for agents with respect to a pair of signals $(\omega_1,\omega_2)\in\{(C,C),(C,D),(D,C),(D,D)\}$. Each components consists of positive $X,Y,Z$ such that $Y<Z$.}
    \label{tab:prof1}
\end{table}
Let $q_0=C$ be an initial state of an agent (or an automaton). For instance, the grim trigger strategy is expressed by 
\begin{align}
    \begin{aligned}
    \delta(C,0)&=C,~~\delta(D,0)=D\\
    \delta(C,1)&=D,~~\delta(D,1)=D
    \end{aligned}
\end{align}
This means that an agent keeps a cooperative mindset while $0$ is observed, but never cooperates once $1$ is observed. Interestingly, strategy profiles which game theorists use are almost the same as state transition diagrams which computer scientists use (Fig.\ref{fig:gt}). 
 Decision making is not a hard task for this automaton and the complexity of this PD is 1 (we define the complexity by the minimal time step needed for an automaton to make a decision "accept" or "reject"). However there is yet another way to introduce a cost function, that is a payoff function. Agents playing a game try to maximize their rewards (or minimize economic loss) by choosing strategies based on opponents' signals. In a single round PD, mutual cooperation $(C,C)$ is not a Nash equilibrium \cite{Nash48}, however there is a chance that the Pareto optimal strategy $(C,C)$ can be an equilibrium of the repeated game. A study on a repeated game is to find such a non-trivial equilibrium solution that can be established in a long term relation of agents. The automaton which decides if a mutual cooperation relation is maintained or not has $Q_\text{acc}=\{C\}$. From this viewpoint of the games/automata correspondence, mixed strategy games can be seen as stochastic automata models. Quantum games which we consider below is a simple version of quantum automata models \eqref{automata}, \eqref{eq:automata} without degrees of spacial freedom.  

\subsection{Repeated Quantum Games}
We apply our model to repeated quantum games.
Historically, quantum games was proposed by \cite{Eisert:1998vv,1999PhRvL..82.1052M} and many relevant works have been made. In many cases, however, only a single round quantum games are studied and less is known for repeated quantum games. Some finite and small repeated quantum games are proposed by \cite{2002PhLA..300..541I,2012JPhA...45h5307F}, whereas infinite cases have not been addressed yet. Especially in terms of repeated games, a study on efficient strategy in a long-term relation is critical. In this section, we formulate the problem and solve it affirmatively. For simplicity we restrict ourselves to two-person prisoners' dilemma where rewards to agents are given in Table. \ref{tab:prof1}. A state of each agent is spanned by two vectors $\ket{C}$ and $\ket{D}$, corresponding to Cooperate and Defect:
\begin{equation}
    \ket{C}=\binom{1}{0},\ket{D}=\binom{0}{1}. 
\end{equation}
We are typically interested in a situation where strategic efficiency is achieved, which we call the Folk theorem. So when agents play the quantum prisoners' dilemma (QPD), we consider what gives incentive for agents to keep $\ket{C}\otimes\ket{C}$ in a long-term relation. In every round, each agent chooses a quantum strategy independently. So a strategic state of an agent $i=1,2$ is generally defined by
\begin{equation}
    \ket{\psi^i_{t}}=S^i_t\ket{0}, 
\end{equation}
where $S^i_t$ should satisfy $\|S^i_t\ket{0}\|^2=1$ for any $t=1,2,\cdots$, called a quantum strategy. Especially we define two special quantum strategies by
\begin{align}
S_{C}&=\ket{C}\bra{C}\\
S_{\overline{C},t}&=a_t\ket{C}\bra{C}+b_t\ket{D}\bra{C},
\end{align}
where $b_t$ is some non-zero complex numbers and $a_t,b_t$ satisfy $|a_t|^2+|b_t|^2=1$. The game proceeds as follows. In each round each gent choose a quantum strategy, operates it to $\ket{0}$. Neither a quantum strategy nor a quantum state is a physical observable, so agents measure quantum states and rewards are paid to agents based on their classical signals $\omega_i\in\{C,D\}$. Information of their outcomes can be either open or close to agents. Such games, where a strategy cannot be directly observed, are often referred to as imperfect monitoring games. If open (closed), the game looks similar to a game with public (private) monitoring. Private monitoring is done by observing their opponents' signals and agents cannot know their own signals.
\begin{table}[]\centering
    \begin{tabular}{c|c|c}
         & C&D \\\hline
        $S_{C,t}$ &1 &0\\
        $S_{\overline{C},t}$ & $|a_t|^2$& $|b_t|^2$
    \end{tabular}
    \caption{Monitoring accuracy with respect to each strategy. }
    \label{tab:prof}
\end{table}

The profit function of agent $i$ is defined by 
\begin{equation}
    \$_i=(1-\delta)\sum_{t=1}^\infty\delta^t\sum_{\omega_1,\omega_2}\$_i(\omega_1,\omega_2)|\bra{\omega_1,\omega_2}S^1_tS^2_t|0,0\rangle|^2,
\end{equation}
where $\delta\in(0,1)$ is a discount factor and $\$_i(\omega_1,\omega_2)$ is $i$'s profit defined in the Table \ref{tab:prof} when a pair $(\omega_1,\omega_2)$ of classical outputs is observed. $\$_i(n)$ is a discrete version of \eqref{eq:aaaa} without degree of freedom about final states. We call $V_i(S^1_t,S^2_t)=\sum_{\omega_1,\omega_2}\$_i(\omega_1,\omega_2)|\bra{\omega_1,\omega_2}S^1_tS^2_t|0,0\rangle|^2$, which is a discrete version of \eqref{eq:v4}, the expected payoff of agent $i$. Regarding the QPD, each player tries to maximise $\$_i$ as much as possible, by choosing $U_t=U^1_t\otimes U^2_t$. 

\begin{thm}[Strategic Efficiency]
There is such a quantum strategy for the repeated QPD that is an equilibrium of the repeated game.   
\end{thm}
More explicitly we can show the following statement. 
\begin{prop}
The (Trigger, Trigger) strategy is an equilibrium of the repeated QPD. 
\end{prop}
Proof is simple. Let $\ket{\psi_0}=\ket{C,C}$ be an initial state of the RQPD. We write $\ket{\overline{C}}=a\ket{C}+b\ket{D}$ with a non-zero $a\in\mathbb{C}$.  Suppose agents play the (Trigger, Trigger) strategy, that is they repeat cooperation $S_C$ until $\ket{D}$ is observed and, once if $\ket{D}$ is observed, they play a not-cooperate strategy $S_{\overline{C}}$. To complete the proof of the statement, it is suffice to show the (Trigger, Trigger) is an equilibrium of the game. Let $r>0$ be $j$'s probability of playing $S_{\overline{C}}$ when $j$ does observe $D$. $r=1$ is called the grim trigger strategy. Suppose both agents play the trigger strategy and cooperative relation is maintained. Then the discounted payoff $\$^*_i$ of agent $i$ is $\$^*_i = V_i(S_{C}, S_{C}) + \delta\$^*_i$, which implies $\$^*_i=\frac{X}{1-\delta}$. By playing $S_{\overline{C},t}$ agent $i$ can increase the expected payoff by $V_i(S_{\overline{C},t},S_{C})-V_i(S_{C},S_{C})=|b_t|^2Y$, but looses the expected rewards in the future $\delta r |b_t|^2\$^*_i$. Hence agent $i$ has no incentive to change the trigger strategy if 
\begin{equation}
\$_i(S_{\overline{C},t},S_{C})-\$_i(S_{C},S_{C})\le \delta r |b_t|^2\$^*_i. 
\end{equation}
Solving this inequality, we obtain the inequality  
\begin{equation}
\delta\ge\frac{Y}{rX+Y}. 
\end{equation}
Therefore (Trigger,Trigger) is an equilibrium strategy, since the R.H.S. is always smaller than 1. 

This result agrees with our naive intuition. Since players know there are no welfare loss while playing $S_C$, they would not be willing to change their strategies unless $r=0$ or $X\ll Y$.  In addition, our observation in the proof above shows that a quantum game is not quiet the same as a classical case. Though quantum games looks similar to imperfect monitoring games with mix-strategy, an imperfect monitoring process usually includes measurement errors and hence triggers welfare loss, which describes a loss of economic efficiency that can occur when equilibrium is not achieved. Therefore (Trigger,Trigger) unlikely becomes an equilibrium of a repetition of prisoners' dilemma with imperfect monitoring \cite{COMPTE2002151}. In contrast to such classical repeated games, our quantum game assures that $\ket{C}$ is observed without any error while $S_C$ is played, therefore welfare loss never occurs. Indeed, this fact makes a clear difference between the CPD and the QPD. From a viewpoint of the conventional classical game theory, if measurement of signals doe not unveil opponents' strategies, which is called conditional independence, "Anti-folk Theorem" claims that only a single round Nash equilibrium can be an equilibrium of the CPD with imperfect private monitoring under conditional independence \cite{MATSUSHIMA1991253}. In contrast to this, we claim that QPD always respect the Folk Theorem, though it also respects conditional independence since agents' quantum states are not entangled at all and measurement does not unveil opponents' strategies.

\if{
\subsection{Remarks}
To this end, we begin with a simple example. Let $\phi=\phi(x,t)$ be a free massive scalar field, which obeys $\partial^2_t\phi=(\nabla^2-m^2)\phi$. We write $\bra{x}=\bra{0}\phi(x,t)$ and a Sch\"{o}dinger picture wavefunction $\psi(x)$ as $\psi(x)=\langle{x}|\psi\rangle$. Since $\partial_t\psi(x)=\langle0|\partial_t\phi(x,t)|\psi\rangle$, the low-energy expression of $\psi(x)$ obeys the Schr\"{o}dinger equation 
\begin{equation}
    i\partial_t\psi(x,t)=-\frac{1}{2m}\partial^2_x\psi(x,t). 
\end{equation}
Then $\psi(x,t)$ respects the equation of continuity \eqref{eq:eoc}. Especially if a planer wave solution $\psi(x,t)=A\exp[i(px-Et)]$ it is easy to see 
\begin{equation}
    \partial_t \mu+\nabla j=0,
\end{equation}
where $\mu=|\psi(x,t)|^2=|A|^2$ and $j=\frac{1}{2mi}(\psi^*\nabla \psi-\psi\nabla\psi^*)=\frac{p}{m}|A|^2$. 
Suppose $U$ is the Feynman kernel  
\begin{align}
\begin{aligned}
    K(y,x:t_1,t_0)&=\int_{x(t_0)=x}^{x(t_1)=y}\mathcal{D}xe^{iS[x(t)]}\\
    &=\bra{y}\exp\left(-i\frac{p^2}{2m}(t_1-t_0)\right)\ket{x}.
\end{aligned}
\end{align}
The term \eqref{H:x} is equivalent to the unitarity. And the term \eqref{H:y} requires $\varphi_1(y)=\psi(y,t_1)$, which is by definition of $K$
\begin{equation}
    \psi(y,t_1)=\int dxK(y,x:t_1,t_0)\psi(x,t_0). 
\end{equation}
The cost is
\begin{equation}\label{eq:cost}
    C(x,y)=\bigg|\int dxdy c(x,y)K(y,x:t_1,t_0)\psi_0(x)\bigg|^2.
\end{equation}
}\fi
\if{Decompose $x(t)$ into the sum $x(t)=x_c(t)+\tilde{x}(t)$ of classical path $x_c(t)$ and the fluctuation $\tilde{x}(t)$, then the action is also decomposed into the sum  $S[x(t)]=S[x_c(t)]+S[\tilde{x}(t)]$ and the Feynman kernel is written as 
\begin{equation}
     K(y,x:t_1,t_0)=f(t_1,t_0)e^{iS[x_c(t)]}, 
\end{equation}
where $f(t_1,t_0)$ is a function of $t_0,t_1$ and is independent on $x,y$. Therefore the amplitude $\psi_1(y)K(y,x:t_1,t_0)\psi_0(x)$ of transporting $\psi_0(x)$ to $\psi_1(y)$ satisfies the condition of the Monge-Kantrovich optimal transportation plan
\begin{align}
    \begin{aligned}
    \int dx |K(y,x:t_1,t_0)\psi_0(x)|^2&=|f(t_1,t_0)|^2|\psi_1(y)|^2\\
    \int dy |K(y,x:t_1,t_0)\psi_0(x)|^2&=|f(t_1,t_0)|^2|\psi_0(x)|^2
    \end{aligned}
\end{align}
Hence up to the time dependent part $f(t_1,t_0)$, the cost \eqref{eq:cost} gives a proper quantum formulation of the optimal transportation. 
}\fi
\if{
We may define cost by
\begin{equation}
   C(x,y)=c(x,y)|\psi_0(x,t_0)\psi_1(y,t_1)|
  ^2, 
\end{equation}
where $\psi_1(y,t_1)$ is defined by
\begin{equation}
    \psi_1(y,t_1)=\int K(y,x;t_1,t_0)\psi_0(x,t_0)dx,
\end{equation}
where the Feynman kernel is defined by the path-integrsal 
\begin{equation}
    K(y,x;t_1,t_0)=\int_{x}^y \mathcal{D}x(t)e^{iS[x(t)]}=\bra{y}\exp\left(-i\frac{p^2}{2m}(t_1-t_0)\right)\ket{x}. 
\end{equation}
$|\psi_0(x,t_0)\psi_1(y,t_1)|^2$ plays a role of $\pi(x,y)$ which are required to satisfy
\begin{equation}\label{eq:munu}
    \int |\psi_0(x,t_0)\psi_1(y,t_1)|^2dx=\nu(y),~\int |\psi_0(x,t_0)\psi_1(y,t_1)|^2dy=\mu(x). 
\end{equation}
Then it is obvious that the conservation of mass law $\int\mu(x)dx=\int\nu(y)dy$ is automatically satisfied. The requirements \eqref{eq:munu} can be implemented as penalty terms of $H$ in such a way that 
\begin{equation}
    H_1=\int dy\left(\int|\psi_0(x)\psi_1(y)|^2dx-\nu(y)\right)^2+\int dx\left(\int|\psi_0(x)\psi_1(y)|^2dy-\mu(x)\right)^2. 
\end{equation}
}\fi

\if{
\section{Quantum Field Theory as a Computation Principle of Nature}
A quantum field computer would be able to efficiently simulate any process that occurs in Nature.
\section{Quantum and Classical Computation Revisit}
classical computer \cite{doi:10.1112/plms/s2-42.1.230} and quantum computer \cite{Feynman1986,doi:10.1098/rspa.1985.0070}

Quantum computation is defined by the following time-evolution of quantum state $\{\ket{\psi_t}\}_{t=1,2,\cdots, T}$, where $T<\infty$ is computation time:
\begin{enumerate}
    \item A quantum computer is always in a state $\ket{\psi_t}$
    \begin{equation}
        \ket{\psi_t}=\sum_{z\in\{0,1\}^n}c_z(t)\ket{z}, 
    \end{equation}
    where $c_z(t)$ is a complex number such that $\sum_{z\in\{0,1\}^n}|c_z(t)|^2=1$. 
    
    \item The quantum computer in an initial state $\ket{\psi_1}$ changes its state, during computation, by a series of unitary operations as 
    \begin{equation}
        U_t: \ket{\psi_t}\to  \ket{\psi_{t+1}}
    \end{equation}
    \item At the end $t=T$ of computation, a state $\ket{z}$ is obtained with probability $|c_z(T)|^2$ by measurement of the quantum computer in a state 
    \begin{equation}
        \ket{\psi_T}=\sum_{z\in\{0,1\}^n}c_z(T)\ket{z}. 
    \end{equation}
\end{enumerate}
A significant change from probabilistic classical computing is that classical computer's state is generally defined by 
\begin{equation}
    \sum_{z\in\{0,1\}^n}c_z(t)\ket{z}, 
    \end{equation}
    where $c_z(t)$ is a non-negative real number such that $\sum_{z\in\{0,1\}^n}c_z(t)=1$. Note that for both classical and quantum computers, the output results are always classical bitstring vector $\ket{z}$ which corresponds one-to-one with $z\in\{0,1\}^n$. 

Quantum annealing \cite{PhysRevE.58.5355} is as powerful as universal quantum computation. Indeed QA can be universal for quantum computation if its dynamics is adiabatic (AQC \cite{Farhi472}) and a given Hamiltonian is a nonstoquastic one \cite{doi:10.1137/080734479}. The computational power of AQC with stoquastic Hamiltonians has been thought to be less powerful than the nonstoquastic cases \cite{RevModPhys.90.015002,2018arXiv180100862P}. However more recently the quantum speedup possibility of AQC with stoquastic Hamiltonians was pursued theoretically \cite{2018arXiv180309954F}, which reported that AQC with an $X$-basis measurement in addition to a $Z$-basis measurement is enough to realize quantum speedup for some problems. 

D-Wave qauntum annealear \cite{Johnson2011}
}\fi


\section*{Acknowledgement}
I thank Travis Humble for useful discussion about quantum computation and game theory. I came up with an idea of showing the Folk theorem while I discussed with him. Also I was benefited by discussing with Katsuya Hashino, Kin-ya Oda and Satoshi Yamaguchi. This work was supported in part by Grant-in-Aid for Scientific Research on Innovative Areas, the Japanese Ministry of Education, Culture, Sports, Science and Technology, No. 19J11073.

\bibliographystyle{JHEP}
\bibliography{ref}
\end{document}